\documentclass[epj]{svjour}
\usepackage{graphics}
\begin{document}
\title{Out of equilibrium quantum field dynamics in external fields}
\subtitle{}
\author{Francisco J. Cao\inst{1,2}
}                     

\institute{Departamento F\'\i sica At\'omica, Molecular y Nuclear.
Universidad Complutense de Madrid. Avenida Complutense s/n, 28040
Madrid. Spain. \and LERMA. Observatoire de Paris, Laboratoire
Associ\'e au CNRS UMR 8112. 61, Avenue de l'Observatoire, 75014
Paris. France.}
\date{Received: date / Revised version: date}
%
\abstract{
   The quantum dynamics of the symmetry broken $ \lambda (\Phi^2)^2 $ scalar field
    theory in the presence of an homogeneous external field is investigated in
    the large $ N $ limit. We consider an initial thermal state of temperature T
    for a constant external field $ \vec {\cal J} $. A subsequent sign flip of the external
    field, $ \vec {\cal J} \to - \vec {\cal J}$, gives rise to an out of equilibrium nonperturbative
    quantum field dynamics. We review here the dynamics for the symmetry
    broken $ \lambda(\Phi^2)^2 $ scalar $N$ component field theory in the large N
    limit, with particular stress in the comparison between the
    results when the initial temperature is zero and when it is
    finite. The presence of a finite temperature modifies the dynamical
    effective potential for the expectation value, and also makes
    that the transition between the two regimes of the early dynamics
    occurs for lower values of the external field. The two regimes
    are characterized by the presence or absence of a temporal trapping
    close to the metastable equilibrium position of the potential.
    In the cases when the trapping occurs it is shorter for larger initial
    temperatures.
} 
\maketitle

\newcommand{\be}{\begin{equation}}
\newcommand{\ee}{\end{equation}}
\newcommand{\bea}{\begin{eqnarray}}
\newcommand{\eea}{\end{eqnarray}}
\newcommand{\bi}{\begin{itemize}}
\newcommand{\ei}{\end{itemize}}
\newcommand{\imply}{\Rightarrow}

\section{Introduction}
\label{intro}

Several important physical systems, as the ultrarelativistic heavy
ion collisions \cite{ioncol} and the early universe \cite{tsuinf},
present out of equilibrium dense concentrations of particles. The
presence of these concentrations imply the need of out of
equilibrium nonperturbative quantum field theory methods, as the
large $N$ limit.

\section{The model}
\label{sec:1}

We consider $N$ scalar fields, $\vec\Phi$, with a $
\lambda(\vec\Phi^2)^2$ selfinteraction in the presence of an
external field $\vec{\cal J}$. The action and the lagrangian
density are given by
\bea
S &=& \int{d^4x {\cal L} }\;, \\
{\cal L} \! &=& \! \frac12[\partial_\mu\vec\Phi(x)]^2 \! - \frac12
m^2 \vec\Phi^2 -\frac{\lambda}{8N}(\vec\Phi^2)^2 \! -
\frac{m^4N}{2\lambda}+ \! \vec{\cal J} \vec\Phi \;\;\;
\label{modelo}
\eea
We restrict ourselves to the case where the symmetry is
spontaneously broken, i.e., $m^2<0$; and we mainly consider small
coupling constants $ \lambda $, because this slows the dynamics
and allows a better study of its different parts.

We consider here the evolution of a initial thermal state of
temperature $T$ after a flip in the homogeneous external field
$\vec{\cal J}\to -\vec{\cal J}$. We can choose the axes in the
$N$-dimensional internal space such that
\be
\vec{\cal J} = \left\{
 \begin{array}{lll}
   (\sqrt{N}J,0,\ldots,0) & \mbox{for} & t\leq 0, \\
   (-\sqrt{N}J,0,\ldots,0) & \mbox{for} &t> 0.
 \end{array}
 \right.
\ee
For an initial thermal state we have an expectation value parallel
to the external field. The following decomposition can be done
\be
\vec\Phi(x) = \left( \sigma(x), \vec\pi(x) \right) =
\left(\sqrt{N} \phi(t) + \chi(x), \; \vec\pi(x) \right)\;,
\ee
with $ \sqrt{N}\phi(t) = \langle \sigma(x) \rangle $; thus,
$\langle \chi(x) \rangle = 0 $. While in the remaining $ N-1 $
directions transversal to the expectation value, $\langle
\vec\pi(x) \rangle=0$.

\section{Evolution equations in the large $ N $ limit}

As we have one direction parallel to the expectation value and
$N-1$ transversal, the fluctuations in the transverse directions
dominate in the large $ N $ limit, while those in the longitudinal
direction only contribute to the evolution equations as
corrections of order $1/N$.
\par
The large $ N $ limit provides explicit evolution equations for
the expectation value and the modes of the quantum fluctuations.
See Refs.~\cite{extfield} and \cite{extfieldT} for the detailed
expressions. We will discuss here the results obtained with this
evolution equations, in particular the effects of the initial
temperature. In order to simplify the expression of the results,
we introduce the following adimensional variables:
\bea
&\tau\equiv|m|t\;,\quad \eta(\tau)\equiv
\sqrt{\frac{\lambda}{2}}\frac{\phi(t)}{|m|}\;,\quad
{\vec \jmath}\equiv\sqrt{\frac{\lambda}{2N}}\frac{{\vec{\cal J}}}{|m|^3}\;,\\
& g\equiv\frac{\lambda}{8\pi^2}\;,\quad
\beta\equiv \frac{\beta_d}{|m|}\;,\quad
g\Sigma(\tau)\equiv\frac{\lambda}{2|m|^2} \frac{\langle \vec\pi^2
\rangle(t)}{N}\;.
\eea

The fact that the initial state is not the ground state but a
thermal state increases the initial value of $ g \Sigma $. It can
be shown that $ \Sigma(0) $ is approximately given by
\be \label{SigmaT}
\Sigma(0) = \Sigma^{T=0}(0) + \frac{\pi^2}{3} \beta^{-2}\;.
\ee
$\Sigma^{T=0}(0)$ is the zero temperature value, that for $g\ll 1$
only depends on $j$ (see Ref.~\cite{extfield}). The second term on
the righthand side of Eq.~(\ref{SigmaT}) is the thermal
contribution computed in the hard thermal loop approximation
\cite{lebellac}. This expression for $ \Sigma(0) $ implies that it
is greater for larger initial temperatures.

\section{Effective dynamical potential for the expectation value}

We can define a potential
\be
V_{de;\tau > 0}(\eta, \Sigma) \equiv \frac{\eta^4}{4} -
\frac{\eta^2}{2} + \frac12\eta^2 g\Sigma - \frac{g\Sigma}{2} +
\frac{(g\Sigma)^2}{4} + \frac14+j\eta\;.
\ee
that can be interpreted as a dynamical effective potential because
the evolution equation for $\eta$ (in the large $N$ limit) can be
written as  \cite{extfield,extfieldT}
\begin{equation}
\ddot\eta(\tau)=-\frac{\partial}{\partial \eta}V_{de;\tau>0}(\eta,
\Sigma)\;.
\end{equation}
It must be stressed that $V_{de;\tau>0}$ is an effective potential
\emph{only} for $\eta$ (and \emph{not} for the modes).

\section{Equilibrium states for the dynamical potential}

The dynamical effective potential for $\tau\leq 0$ can be defined
as $V_{de;\tau\leq 0}(\eta, \Sigma) = V_{de;\tau>0}(\eta,\Sigma)
-2j\eta$. Therefore, the stationary states for the initial
dynamics for times $\tau\leq 0$ (before the external field has
been flipped) are the solutions of $V_{de;\tau\leq 0}'(\eta) = 0$
(the prime means $\eta$ derivative). Thus, they verify
\be \label{statstates}
\eta^3+(-1+g\Sigma)\eta -j =0\;.
\ee
For small external fields,
\be \label{jd}
j<j_d \equiv 2\sqrt{\frac{(1-g\Sigma)}{27}},
\ee
we have three roots. There is a global minimum that corresponds to
a stable equilibrium state, a local minimum that corresponds to a
metastable equilibrium state, and a local maximum (unstable
equilibrium) (see Fig.~\ref{potenciales1}). On the other hand, for
larger external fields ($ j > j_d $) there is a single extreme
that corresponds to a global minimum (stable equilibrium).

As a larger initial temperature increases $ g\Sigma(0) $, it
contributes to restoring the symmetry and as consequence makes the
value of $ j_d $ lower [see Eq.~(\ref{jd})].

We consider here the more interesting case $ j < j_d $, where a
potential barrier is present.

\section{Early time dynamics and dynamical regimes} \label{sec:2}

\begin{figure}
\resizebox{0.5\textwidth}{!}{%
  \includegraphics{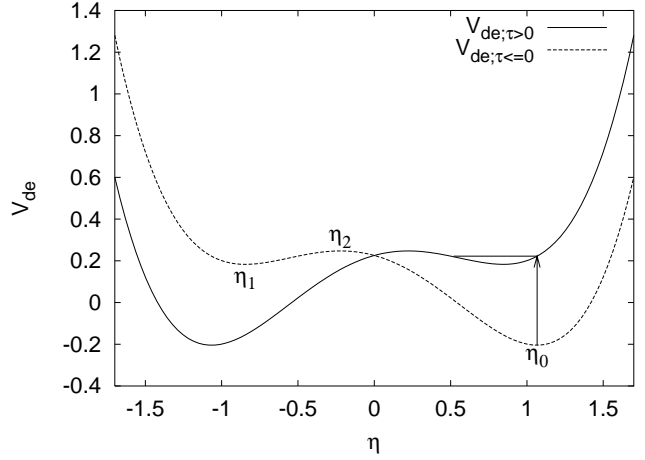}
}
\caption{Dynamical effective potential $V_{de;\tau\leq 0}$ and
$V_{de;\tau> 0}$ as a function of $\eta$ for $g\Sigma(0)=0.05$.
The value of the external field is $j=0.2$. The positions of the
global minimum $\eta_0$, the local minimum $\eta_1$, and the local
maximum $\eta_2$ are shown. There is a potential barrier ($ j <
j_d $), and the systems gets temporally trapped ($ j < j_c$).
\label{potenciales1}}
\end{figure}

\begin{figure}
\resizebox{0.5\textwidth}{!}{%
  \includegraphics{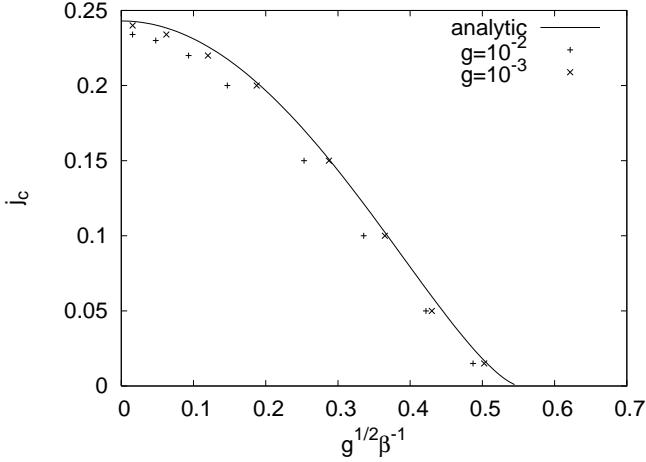}
} \caption{Critical external field $j_c$ as a function of
$\beta^{-1}$ for $g=10^{-2}$ and $ g=10^{-3}$ obtained from
numerical simulations (symbols) and from the analytical formula in
Eq.~(\ref{jcbeta}) (full line). There are two regions in the
$(j,\beta^{-1})$ plane corresponding to the two dynamical regimes:
for $j<j_c$ the system is temporally trapped in the metastable
state; for $j>j_c$ the system rapidly reaches the neighborhood of
the global minimum of the potential. \label{figjcbeta}}
\end{figure}

After the initial flip of the external field sign at $ \tau=0 $
the positions of the absolute minimum and the relative minimum of
the potential are interchanged, and the state of the system
becomes a metastable state (Fig.~\ref{potenciales1}). The
existence of a potential barrier gives rise to two different
dynamical regimes. In the first one the system can directly
overcome the barrier, and rapidly reaches the neighborhoods of the
global minimum. While in the second regime the system can not
overcome the barrier directly, and it gets temporally trapped
close to the metastable state. The external field critical value
$j_c$ that separates the two regimes, untrapped $|j|>j_c$, or
trapped $|j|<j_c$, is
\be \label{jcbeta}
j_c(\beta) = \frac{j_c^{T=0}}{\left[1-
\left(\frac{\beta_c^{J=0}}{\beta}\right)^{2}\right]^{3/2}}
\ee
with $ j_c^{T=0} = \sqrt{ 2 \; {\left( 13^2 + 15 \sqrt{5} \right)
\over 19^3}} = 0.243019\ldots \;, $ and $
\beta_c^{J=0} = \pi\sqrt{\frac{g}{3}}. $
Therefore, we see that if the initial temperature is higher enough
the system will not be trapped. In addition, Eq.~(\ref{jcbeta})
states how the trapping can disappear due to a combination of the
effects of both the initial temperature and the external field (see fig. \ref{figjcbeta}).

\par
When the trapping is present spinodal instabilities made the modes
grow and finally their backreaction allows the system to go close
to the stable state. A detailed analysis of the spinodal
instability allows to obtain the trapping time, or spinodal time
$\tau_s$ (see Refs.~\cite{extfield,extfieldT}). At early times ($
\tau < \tau_s $), the effective squared mass oscillates with a
negative average, $ - \mu^2 \simeq - j $ implying the growth of
the modes due to spinodal instability. This implies a
quasiexponential growth of $ g\Sigma(\tau) $ for $ \tau < \tau_s $
\begin{equation}  \label{gSigmaTmain}
g\Sigma_s(\tau)\approx \left\{
 \begin{array}{lcc}
   \frac{2}{\beta\mu}g\Sigma_s^{T=0}(\tau) & \mbox{for} &
      \beta^{-1} \gg \mu \;, \vspace{2mm}\\
    g\Sigma_s^{T=0}(\tau) & \mbox{for} &
      \beta^{-1} \ll \mu \;,
 \end{array}
 \right.
\end{equation}
with
\be \label{gSigmaT0main}
g\Sigma_s^{T=0}(\tau)\approx\frac{g\sqrt{\pi\mu}e^{2\tau\mu}}{8\tau^{3/2}}\;
\ee
the value for zero temperature. After a certain time, the spinodal
time $\tau_s$, the quantum and thermal effects start to be
important in the dynamics, $g\Sigma_s(\tau_s)$ compensates
$-\mu^2$ and the exponential growth of the mode functions stops,
then the mode functions start to have an oscillatory behavior.
Thus, the spinodal time $\tau_s$ is defined as
\begin{equation} \label{tsmain}
g\Sigma_s(\tau_s)=\mu^2\;.
\end{equation}
A good approximate expression for the spinodal time is
\be \label{tsfinal}
\tau_s \! = \! \left\{
  \begin{array}{lcc}
    \frac{1}{2\mu} \left[\log\left(\frac{8}{g\sqrt{\pi}}\right)
    - \log\left(\frac{2}{\beta\mu}\right) + \frac{3}{2}
    \log(\mu\tau_s) \right]
    & \!\! \mbox{ for } &  \!\! \beta^{-1} \!\! \gg \! \mu \vspace{1 mm}\\
    \frac{1}{2\mu} \left[\log\left(\frac{8}{g\sqrt{\pi}}\right)
    + \frac{3}{2} \log(\mu\tau_s) \right]
    & \!\! \mbox{for} &  \!\! \beta^{-1} \!\! \ll \! \mu \\
  \end{array}
\right.
\ee
For $\beta^{-1} \gg \mu$ we see that a higher initial temperature
implies a shorter trapping period.

\section{Intermediate time dynamics}

After the early dynamics described in the previous section, the
system enters a quasiperiodic regime, both for $ j<j_c $ and for $
j>j_c $. The system at this intermediate times presents a clear
separation between fast variables and slow variables. $ \eta(\tau)
$, $ g\Sigma(\tau) $, and the effective squared mass oscillate
fast, while the amplitude of their oscillations slowly decreases.
This quasiperiodic regime in the large $N$ limit evolution
equations is present both for zero and nonzero initial
temperatures. (See Refs.~\cite{extfield,extfieldT}.)

\section{Conclusions}

One of the main features of the dynamics is the existence in some
cases of a trapping stage in the dynamics, that have been
characterized. In particular, we have seen that the main influence
of the initial temperature is to contribute to the restoration of
the symmetry of the potential, and as a consequence a higher
initial temperature shortens or even avoids the initial trapping.
For intermediate times we have found a quasiperiodic regime.
However, how much this dynamics will be modified by the inclusion
of next to leading order terms in the large $N$ approximation
\cite{berges} is still an open question.

\section{Acknowledgments}

We thank H. J. de Vega and M. Feito for fruitful collaborations in 
this problem. We acknowledge financial support from the
Ministerio de Educaci\'on y Ciencia through Research Projects Nos.
BFM2003-02547/FISI and FIS2006-05895.

\end{document}